\documentclass{aastex631}

\usepackage{bm}
\usepackage{subcaption}
\usepackage[T1]{fontenc}
\usepackage{booktabs}
\usepackage{threeparttable}
\usepackage{multirow}
\usepackage{amsmath}
\usepackage[utf8]{inputenc}
\usepackage{makecell}
\usepackage{hyperref}
\usepackage{numprint}
\usepackage{rotating} 

\usepackage{tabularx}

\begin{document}

\title{Synchronization in Traffic Dynamics: Mechanisms of Hysteresis}
 
\author{Jinghui Wang}
\altaffiliation{{\url{zhux083@gmail.com}}}
\affiliation{School of Safety Science and Emergency Management\\
Wuhan University of Technology\\
Wuhan, China}

\author{Wei Lv}
\altaffiliation{{\url{weil@whut.edu.cn}}}

\affiliation{School of Safety Science and Emergency Management\\
Wuhan University of Technology\\
Wuhan, China}
\affiliation{China Research Center for Emergency Management\\
Wuhan University of Technology\\
Wuhan, China}

\begin{abstract}

Starting from a second-order linear differential equation, we analyze the dynamical mechanisms of no behavior pattern (pure response), reaction and anticipation behaviors in traffic. As an emergence of the underlying dynamical evolution, the periodic evolution trajectories (3D hysteresis) in phase space ($v_i, v_j, d_{ji}$) exhibit fascinating characters. We investigate the emerging Time-Delay ($TD$) phenomena and the resulting analytical hysteresis, an equal frequency sets of Lissajous figures. By quantifying energy dissipation through individual and system perspectives, we demonstrate that $TD$ and Time-To-Collision ($TTC$) are direct metrics of zero-dissipation under equilibrium and synchronization states. Finally, a phase diagram based on $TD$ and $TTC$ is developed to bridge the dynamical behaviors in traffic across $\mathbb{R}^1$ and $\mathbb{R}^2$ spaces. Our results provide a theoretical foundation upon which many obscure mechanisms become self-evident, such as the $TD$-induced flip of the hysteresis (clockwise to counterclockwise in FD) and the crossed hysteresis, etc. 

\end{abstract}


\section{Introduction}

The phenomenon of hysteresis in traffic has long received extensive attention \citep{Treiterer}. Researchers have understood that the traffic hysteresis originates from the Time-Delay ($TD$) phenomena in dynamics, based on response mechanisms \citep{zhang1999mathematical}. The reaction and anticipation behavior in the dynamical evolution \citep{treiber2006delays, kesting2008reaction, cordes2023single} will result in the $TD$ variation and manifested as the expansion and contraction of the hysteresis. Established from numerical and analytical investigation, researchers have analyzed many aspects of the hysteresis, including the evolution orientation of hysteresis  \citep{laval2011hysteresis,gayah2011clockwise,jiang2024dynamic}, the connection between the hysteresis and string stability \citep{mattas2023relationship}, hysteresis with multiple rendezvous points \citep{zhang1999mathematical,ni2025there}, and the performance of hysteresis in the relative space-speed phase plane \citep{ni2025there}, among others. A fundamental problem is that these properties appear to be interconnected yet lack a unified theoretical explanation. To this end, we attempt to employ a concise framework to bridge these concepts.

\section{Dynamic Framework}

In this study, the dynamical equations are developed based on a fundamental characteristic of traffic: the constant time headway ($t_h$) strategy. We set $t_h=1.3$ s, which is a common empirical character in pedestrian context. Within 1D system of particles, the desired speed (optimal velocity) of traffic unit $i$ under constrained motion ($0 < v_i < v_{\text{max}}$) for a given headway is defined as $v_d = (x_j - x_i) / t_h$, where $x_j$ represents the position of the preceding traffic unit $j$. The dynamic response of the follower traffic unit $i$ as it approaches its desired speed can be described by the equation $a = (v_d - v_i) / \tau$, where $\tau$ denotes the relaxation time and set to 0.5 s. Following these settings, when the speed ranges between $0$ and $v_{\text{max}}$, the dynamical system can be complete expressed as a second order linear differential equation. Let $\delta$ represent the corresponding time scale of the behavior patterns, where $\delta < 0$ signifies reaction behavior and $\delta > 0$ signifies anticipation behavior. Based on these definitions, dynamical equations for different behavioral patterns can be derived, and ensuring that all parameters maintain explicit physical meanings. In the following, four specific models will be introduced.

\subsection{OV Model (No Behavior Pattern, Pure Response Mechanism)}

When behavioral patterns are not considered, traffic evolution is dominated by the pure response, where the desired speed is defined as $v_{d}=\frac{x_j-x_i}{t_h}$. The underlying dynamics are governed by the differential equation, i.e., linearized OV model \citep{bando1995dynamical}:

\begin{equation}
    \ddot{x}_i + \frac{\dot{x}_i}{\tau} = \frac{x_j - x_i}{t_h \tau}
\end{equation}

Where $\tau$ and $t_h$ are positive time constants. This study focuses on the oscillation dynamics surrounding the synchronization state. Consequently, in a dynamical system governed by second order linear differential equations, we can shift the oscillations of the synchronized traffic flow to the origin within the phase space. This allows for the construction of dynamical equations specifically regarding the perturbation terms:

\begin{equation}
\ddot{\xi}_i + \frac{\dot{\xi}_i}{\tau} = \frac{\xi_j - \xi_i}{t_h \tau}.
\label{xi_ov}
\end{equation}

Based on the properties of linear differential equations, for traffic speeds within the range $v \in (0, v_\text{max})$, the governing equation of the oscillation describes the complete evolution of the traffic flow as it deviates the synchronized state. Clearly, Eq.~\eqref{xi_ov} represents a response unit. To obtain the analytical dynamic results, the Laplace transform is applied to this governing equation, yielding the following poles from the characteristic equation:

\begin{equation}
    s_{1,2} = -\frac{1}{2\tau} \pm \frac{1}{\sqrt{t_h\tau}} \sqrt{\frac{t_h}{4\tau} - 1}.
\end{equation}

Based on the parameter settings where $t_h = 1.3$ s and $\tau = 0.5$ s, it can be determined that the poles are complex conjugate pairs with negative real parts, and the transient response exhibits an underdamped pattern. By applying the direct input ${\xi}_j$, taking the Laplace transform under zero initial conditions, and solving for $G_\text{ov}(s) = {\xi}_i(s)/{\xi}_j(s)$, we obtain:

\begin{equation}
    G_\text{ov}(s) = \frac{1}{t_h \tau s^2 + t_h s + 1}.
    \label{eq:tf1}
\end{equation}

Given a wave function input ${\xi}_j(t) = A \sin(\omega t)$ representing the simple harmonic oscillation of position, corresponding steady state output can be derived based on the transfer function as follows:

\begin{equation}
    {\xi}_{i}(t) = A \cdot |G_{\text{ov}}(j\omega)| \cdot \sin(\omega t + \phi_{\text{ov}}),
    \label{eq:generic_ss_output}
\end{equation}

Where $G_{\text{ov}}(s)$ is the specific system transfer function and $\phi_{\text{ov}} = \angle G_{\text{ov}}(j\omega)$. Let $M_{\text{ov}} =|G_{\text{ov}}(j\omega)|$, we obtain the steady-state response wave:

\begin{equation}
    {\xi}_{i,\text{ov}}(t) = A \cdot M_\text{ov} \cdot \sin(\omega t + \phi_\text{ov}),
\end{equation}

with

\begin{equation}
\begin{cases}
    M_\text{ov} &= \frac{1}{\sqrt{(1 - t_h\tau\omega^2)^2 + (t_h\omega)^2}}, \\
    \phi_{\text{ov}} &= - \arg\bigl(1 - t_h\tau\omega^2 + j t_h\omega\bigr).
    \label{eq:phase1}
\end{cases}
\end{equation}

Now, we have obtained the analytical wave output of the response system under the given input. By extending these results to the complex plane, we can analytically examine all properties of this dynamical system.

\subsection{Reaction Model (Reaction Behavior + Response Mechanism)}

The reaction behavior pattern, characterized by a systematic lag of the input in the response unit, can be described using delay-differential equations (DDEs). In the dynamic equation, we define the time scale of the reaction behavior as $\delta$, which yields the desired speed $v_{d} = \frac{x_{j}(t+\delta) - x_{i}(t)}{t_{h}}, \delta < 0$. The corresponding DDE is:

\begin{equation}
\ddot{x}_i + \frac{\dot{x}_i}{\tau} = \frac{x_{j(t+\delta)} - x_i}{t_h \tau}.
\end{equation}

In line with the previous analysis, we obtain the dynamic equation of the perturbation term as it deviates from the synchronized flow:

\begin{equation}
\ddot{\xi}_i + \frac{\dot{\xi}_i}{\tau} = \frac{{\xi}_{j(t+\delta)} - \xi_i}{t_h \tau}.
\end{equation}

The poles of this dynamical system are consistent with those of the OV model, meaning they share the same transient response pattern. Taking the Laplace transform under zero initial conditions and solving for $G_\text{re}(s) = {\xi}_i(s)/{\xi}_j(s)$ yields:

\begin{equation}
    G_\text{re}(s) = \frac{e^{\delta s}}{t_h \tau s^2 + t_h s + 1}.
    \label{eq:tf3}
\end{equation}

Given the wave function input ${\xi}_j(t) = A \sin(\omega t)$, the steady state wave response is given by:

\begin{equation}
    {\xi}_{i,\text{re}}(t) = A \cdot M_\text{re} \cdot \sin(\omega t + \phi_\text{re}),
\end{equation}
with

\begin{equation}
\begin{cases}
    M_\text{re} &= \frac{1}{\sqrt{(1 - t_h\tau\omega^2)^2 + (t_h\omega)^2}}, \\
    \phi_\text{re} &= \delta\omega - \arg\bigl(1 - t_h\tau\omega^2 + j t_h\omega\bigr).
    \label{eq:phase3}
\end{cases}
\end{equation}

\subsection{CosForce Model (Anticipation Behavior + Response Mechanism)}

The anticipation behavior pattern is characterized by the response unit estimating future states based on a given input to establish the response dynamics. The desired speed is defined as $v_{d} = \frac{\hat{x}_{j}(t+\delta) - x_{i}}{t_{h}}, \delta > 0$. Here, $\hat{x}_{j}(t+\delta) = x_{j} + \delta \dot{x}_{j}$ represents the first order Taylor expansion estimator of $x_{j}$ at time $t+\delta$. We refer to this as the CosForce model, and the governing dynamic equation is:

\begin{equation}
\ddot{x}_i + \frac{\dot{x}_i}{\tau} = \frac{(x_j + \delta \dot{x}_j) - x_i}{t_h \tau}.
\end{equation}

Leads to the modified perturbation dynamics:

\begin{equation}
\ddot{\xi}_i + \frac{\dot{\xi}_i}{\tau} = \frac{({\xi}_j + \delta \dot{{\xi}}_j) - \xi_i}{t_h \tau}.
\end{equation}

The form of the homogeneous differential equation remains unchanged, leading to the identical poles of these dynamical systems. By taking the Laplace transform under zero initial conditions and solving for $G_\text{cf}(s) = {\xi}_i(s)/{\xi}_j(s)$, we obtain:

\begin{equation}
    G_\text{cf}(s) = \frac{1 + \delta s}{t_h \tau s^2 + t_h s + 1}.
    \label{eq:tf2}
\end{equation}

Under the wave function input ${\xi}_j(t) = A \sin(\omega t)$, the steady state wave response is given by:

\begin{equation}
    {\xi}_{i,\text{cf}}(t) = A \cdot M_\text{cf} \cdot \sin(\omega t + \phi_\text{cf}),
\end{equation}

with

\begin{equation}
\begin{cases}
    M_\text{cf} &= \frac{\sqrt{1 + (\delta\omega)^2}}{\sqrt{(1 - t_h\tau\omega^2)^2 + (t_h\omega)^2}}, \\
    \phi_\text{cf} &= \arg\bigl(1 + j\,\delta\omega\bigr) - \arg\bigl(1 - t_h\tau\omega^2 + j\,t_h\omega\bigr).
    \label{eq:phase2}
\end{cases}
\end{equation}

\subsection{FVD Model (Anticipation Behavior + Response Mechanism)}

Another classic form of the anticipation behavior model defines the desired speed as $v_{d} = \frac{\hat{x}_{j}(t+\delta) - \hat{x}_{i}(t+\delta)}{t_{h}}$, $\delta > 0$. Here, $\hat{x}_{j}(t+\delta) - \hat{x}_{i}(t+\delta) = x_{j} + \delta \dot{x}_{j} - x_{i} - \delta \dot{x}_{i}$ represents the first order Taylor expansion estimator of the relative distance $x_{j} - x_{i}$ at time $t+\delta$. This model corresponds to the linearized FVD model \citep{jiang2001full}:

\begin{equation}
    \ddot{x}_i + \frac{\dot{x}_i}{\tau}
    = \frac{(x_j - x_i) + \delta(\dot{x}_j - \dot{x}_i)}{t_h \tau}.
\end{equation}

Leads to the perturbation dynamics
\begin{equation}
    \ddot{{\xi}}_i + \frac{\dot{\xi}_i}{\tau}
    = \frac{({\xi}_j - {\xi}_i) + \delta(\dot{\xi}_j - \dot{\xi}_i)}{t_h \tau}.
    \label{eq:fvd_ode}
\end{equation}

Apply the Laplace transform, the poles of the FVD model are

\begin{equation}
    s_{1,2} = \frac{-(t_h + \delta) \pm
    \sqrt{(t_h + \delta)^2 - 4 t_h \tau}}{2 t_h \tau}.
\end{equation}

The additional term $\delta > 0$ shifts both poles further to the left in the complex plane, resulting in enhanced damping within the FVD model. The associated transfer function is:

\begin{equation}
    G_\text{fvd}(s)
    = \frac{1 + \delta s}{t_h \tau s^2 + (t_h + \delta)s + 1}.
    \label{eq:tf_fvd}
\end{equation}

Given the wave function input ${\xi}_j(t) = A \sin(\omega t)$, the steady state response is:

\begin{equation}
    {\xi}_{i,\text{fvd}}(t)
    = A \cdot M_\text{fvd} \cdot
      \sin\left(\omega t + \phi_\text{fvd}\right),
\end{equation}

where

\begin{equation}
\begin{cases}
    M_\text{fvd} = \frac{\sqrt{1 + (\delta\omega)^2}} {\sqrt{(1 - t_h\tau\omega^2)^2 +\bigl((t_h + \delta)\omega\bigr)^2}}, \\
    \phi_\text{fvd} = \arg\bigl(1 + j\,\delta\omega\bigr) - \arg\bigl(1 - t_h\tau\omega^2 + j\,(t_h + \delta )\omega\bigr).
\end{cases}
\end{equation}

\subsection{Stability of String Dynamics}

\begin{figure}[ht!]
\nolinenumbers
\centering
\includegraphics[scale=0.65]{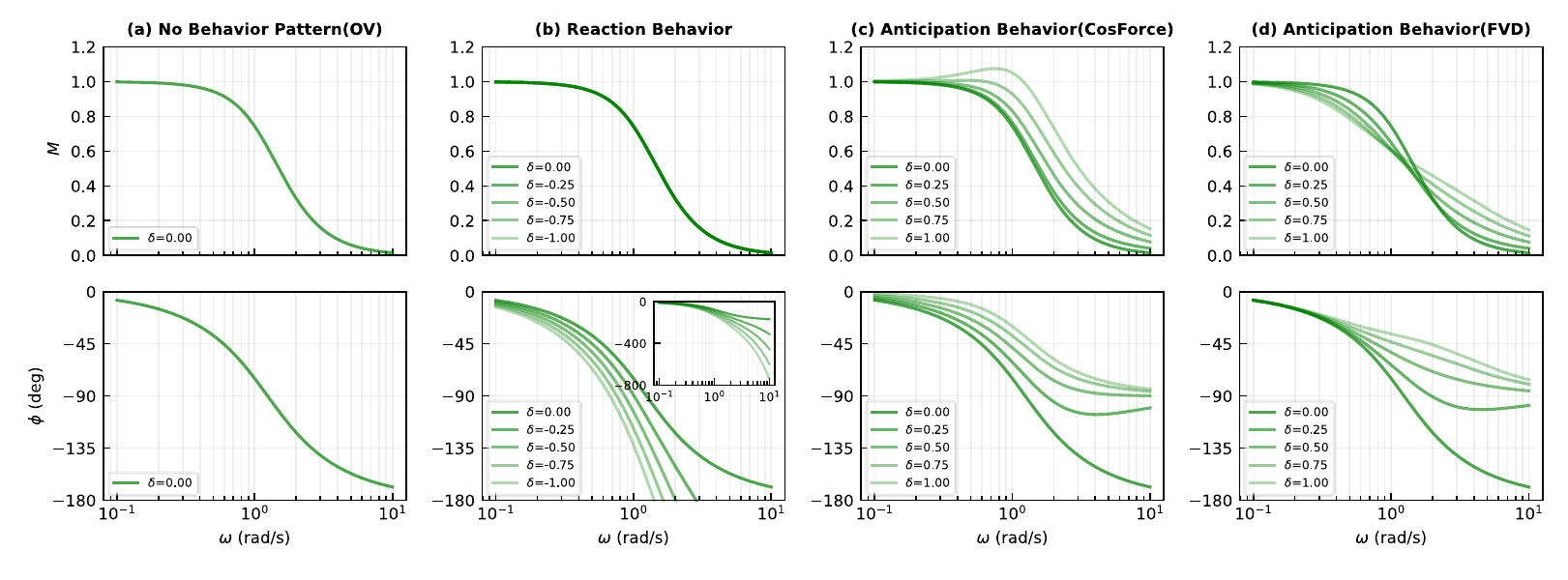}
\caption{Bode diagram of transfer function ($G(j\omega)$) in the four models.}
\label{fig1}
\end{figure}

Based on the above analysis of complex frequency domain, we obtain the analytical forms for the steady state amplitude gain $M$ and phase $\phi$. Corresponding analytical Bode plot is presented in Fig.~\ref{fig1}. According to the results, it can be observed that the dynamical system under discussion exhibits low pass characteristics. From the magnitude plot, we can evaluate the string stability of the traffic, where the stability criterion is defined by $M \le 1$. In the OV model, according to Eq.~\eqref{eq:phase1}, corresponding to stability condition is $t_h \ge  2\tau, \forall \omega>0$. Similarly, we can conclude that the string stability condition of the reaction model is identical to that of the OV model, owing to their perfectly consistent magnitude plot characteristics. The corresponding stability condition is $t_h \ge 2\tau$ for all $\omega > 0$.

\begin{figure}[ht!]
\nolinenumbers
\centering
\includegraphics[scale=0.6]{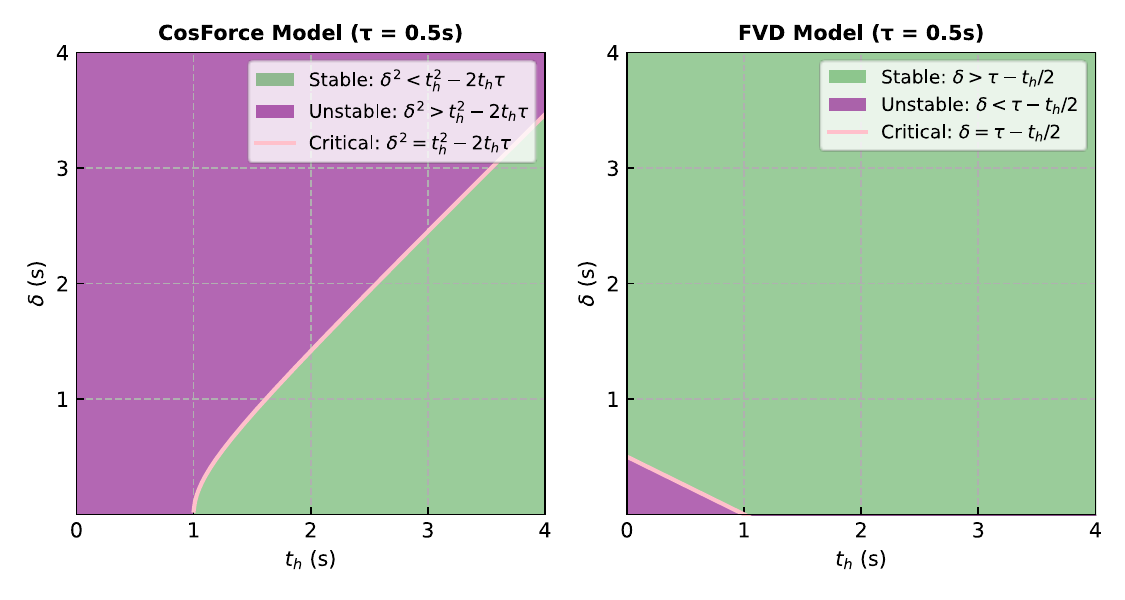}
\caption{Phase diagrams for string stability in CosForce and FVD models.}
\label{fig2}
\end{figure}

In the anticipation model, the introduction of the time scale $\delta$ clearly modifies the form of the transfer function, thereby leading to changes in string stability. In the CosForce model, the stability condition for string dynamics is $M_{\text{cf}} \le 1$, which corresponds to the condition $\delta^2 \le t_h^2 - 2t_h\tau, \forall \omega > 0$. This appears to be a quite strict condition in traffic. Conversely, in the FVD model, the stability condition for string dynamics is $M_{\text{fvd}} \le 1$, corresponding to the stability condition $\delta \ge \tau - t_h/2, \forall \omega > 0$. Fig.~\ref{fig2} presents the stability phase diagrams for the two anticipation models under given parameters, where $t_h = 1.3 \text{ s}$ and $\tau = 0.5 \text{ s}$ are constant parameters. We can clearly observe that the stability conditions of the FVD model are more relaxed. This indicates that the FVD model offers greater flexibility in parameter settings in traffic platoon modeling.

\section{Analytical Hysteresis}

Based on the analytical wave function derived, we can analyze various properties of the traffic oscillations, the most important is hysteresis. From the phase plot in Fig.~\ref{fig1}, we observe that the output of the dynamical system exhibits a phase lag relative to the input. What statistical consequences will such a lag produce? How does this lag influence the geometry of hysteresis? This section will discuss these properties. Before we dive into the analysis, it should be clarified that for the sake of simplicity, our oscillation analysis essentially shifts the system from the synchronization point to the origin. When the conversion from relative states to absolute states is required, the process is essentially just an inverse translation operation.

\subsection{Time-Delay}

When transforming the wave function from the complex frequency domain to the time domain, the phase lag leads to a systematic $TD$. This is also a significant empirical mechanism in traffic statistics. In the dynamical system of this study, the analytical results based on $TD$ can help us understand the factors that influence the hysteresis.

\subsubsection{Wave Functions}

We have defined a input wave function ${\xi}_j(t) = A \sin(\omega t)$ representing the position perturbation of leader unit $j$. The resulting steady state response of follower unit $i$ is ${\xi}_i(t) = A \cdot M \cdot \sin\left(\omega t + \phi\right)$. From this, it can be derived that the speed oscillation $\dot{{\xi}}_j$ of the leader unit $j$ is

\begin{align}
    \dot{{\xi}}_j(t) = A\omega \cos(\omega t).
\end{align}

Corresponding amplitude of $\dot{{\xi}}_j$ is $A_{\dot{\xi}_j} = A \omega$.

The speed oscillation $\dot{{\xi}}_i$ of follower unit $i$ is

\begin{align}
    \dot{{\xi}}_i(t) = A M\, \omega
       \cos\bigl(\omega t + \phi\bigr).
\end{align}

Corresponding amplitude of $\dot{\xi}_i$ is 
$A_{\dot{\xi}_i} = A M \omega$.

The oscillation of relative distance ${\xi}_{ji}$ between $j$ and $i$ is: 

\begin{align}
    {\xi}_{ji}(t) = {\xi}_j(t) - {\xi}_{i}(t) = A \sin(\omega t)
    - A M\, \sin\bigl(\omega t + \phi\bigr).
\end{align}

Corresponding amplitude of $\xi_{ji}$ is $A_{\xi_{ji}} = A \sqrt{1 + M^2 - 2M \cos\phi}$.

The oscillation of relative velocity $\dot{{\xi}}_{ij}$ between $i$ and $j$ is
\begin{align}
    \dot{{\xi}}_{ij}(t) =\dot{{\xi}}_{i}(t) - \dot{{\xi}}_j(t)
    = A M\,\omega \cos\bigl(\omega t + \phi\bigr) -  A\omega \cos(\omega t).
\end{align}

Corresponding amplitude of $\dot{{\xi}}_{ij}$ is $ A_{\dot{{\xi}}_{ij}}=A\omega \sqrt{1 + M^2 - 2M \cos\phi}$.

\subsubsection{Phase Relationships}

After obtaining the wave functions for each state variable $(\dot{{\xi}}_{j},\dot{{\xi}}_{i}, {\xi}_{ji}, \dot{\xi}_{ij})$, we are able to analyze the phase relationships. The phase lag between the leader speed oscillation $\dot{{\xi}}_{j}(t)$ and the relative distance oscillation ${\xi}_{ji}(t)$ is:

\begin{equation}
    \phi_{\dot{\xi}_j, \xi_{ji}} = \phi_{\dot{\xi}_j} - \phi_{\xi_{ji}} = \frac{\pi}{2} - \arg\left(1 - Me^{j\phi}\right)
\end{equation}

By shifting the oscillation part to the DC component of synchronization state, we can find that the $TD$ between the leader speed $v_{j}(t)$ and the relative distance $d_{ji}(t)$ is $TD_{v_j, d_{ji}}=TD_{\dot{\xi}_j, \xi_{ji}}=\phi_{\dot{\xi}_j, \xi_{ji}}/\omega$. 

The phase lag between relative distance oscillation ${\xi}_{ji}(t)$ and follower speed oscillation $\dot{{\xi}}_i(t)$ is:

\begin{equation}
    \phi_{\xi_{ji}, \dot{\xi}_i} = \phi_{\xi_{ji}} - \phi_{\dot{\xi}_i} = \arg\,\left(1 - Me^{j\phi}\right) - \phi - \frac{\pi}{2}.
\end{equation}

Therefore, we can obtain that the $TD$ between the relative distance ${d}_{ji}(t)$ and the follower speed ${v}_i(t)$ is $TD_{d_{ji}, v_i}=TD_{\xi_{ji}, \dot{\xi}_i}=\phi_{\xi_{ji}, \dot{\xi}_i}/\omega$.

The phase lag between leader speed oscillation $\dot{{\xi}}_j(t)$ and follower speed oscillation $\dot{{\xi}}_i(t)$ is:

\begin{equation}
    \phi_{\dot{\xi}_j, \dot{\xi}_i} = \phi_{\dot{\xi}_{j}} - \phi_{\dot{\xi}_i}= -\phi = \arg\bigl(1 - t_h\tau\omega^2 + j t_h\omega\bigr).
\end{equation}

Correspondingly, the $TD$ between the leader speed $v_j(t)$ and the follower speed $v_i(t)$ is given by $TD_{v_j, v_i}=TD_{\dot{\xi}_j, \dot{\xi}_i}=\phi_{\dot{\xi}_j, \dot{\xi}_i}/\omega$. 

The phase lag between relative distance oscillation ${\xi}_{ji}(t)$ and relative speed oscillation $\dot{{\xi}}_{ij}(t)$ is:

\begin{equation} 
\phi_{\xi_{ji}, \dot{\xi}_{ij}} = \phi_{\xi_{ji}} - \phi_{\dot{\xi}_{ij}} = \arg\left(1 - Me^{j\phi}\right) - \arg\left[-j\omega\left( 1 - Me^{j\phi}\right)\right]= \frac{\pi}{2}. 
\label{phi}
\end{equation}

Corresponding, The $TD$ between the relative distance $d_{ji}(t)$ and the relative speed $v_{ij}(t)$ is given by $TD_{d_{ji}, v_{ij}}=TD_{\xi_{ji}, \dot{\xi}_{ij}}=\phi_{\xi_{ji}, \dot{\xi}_{ij}}/\omega$.

In this analysis, the $TD$ between relative distance $d_{ji}(t)$ and follower speed $v_i(t)$ is crucial, as it captures the flip of hysteresis in the $({\xi}_{ji}, \dot{\xi}_i)$ plane, which will be discussed in the following section. Hereafter, $TD$ refers specifically to $TD_{d_{ji}, v_i}$ in the subsequent sections.

\subsection{Hysteresis}

\begin{figure}[ht!]
\nolinenumbers
\centering
\includegraphics[scale=0.65]{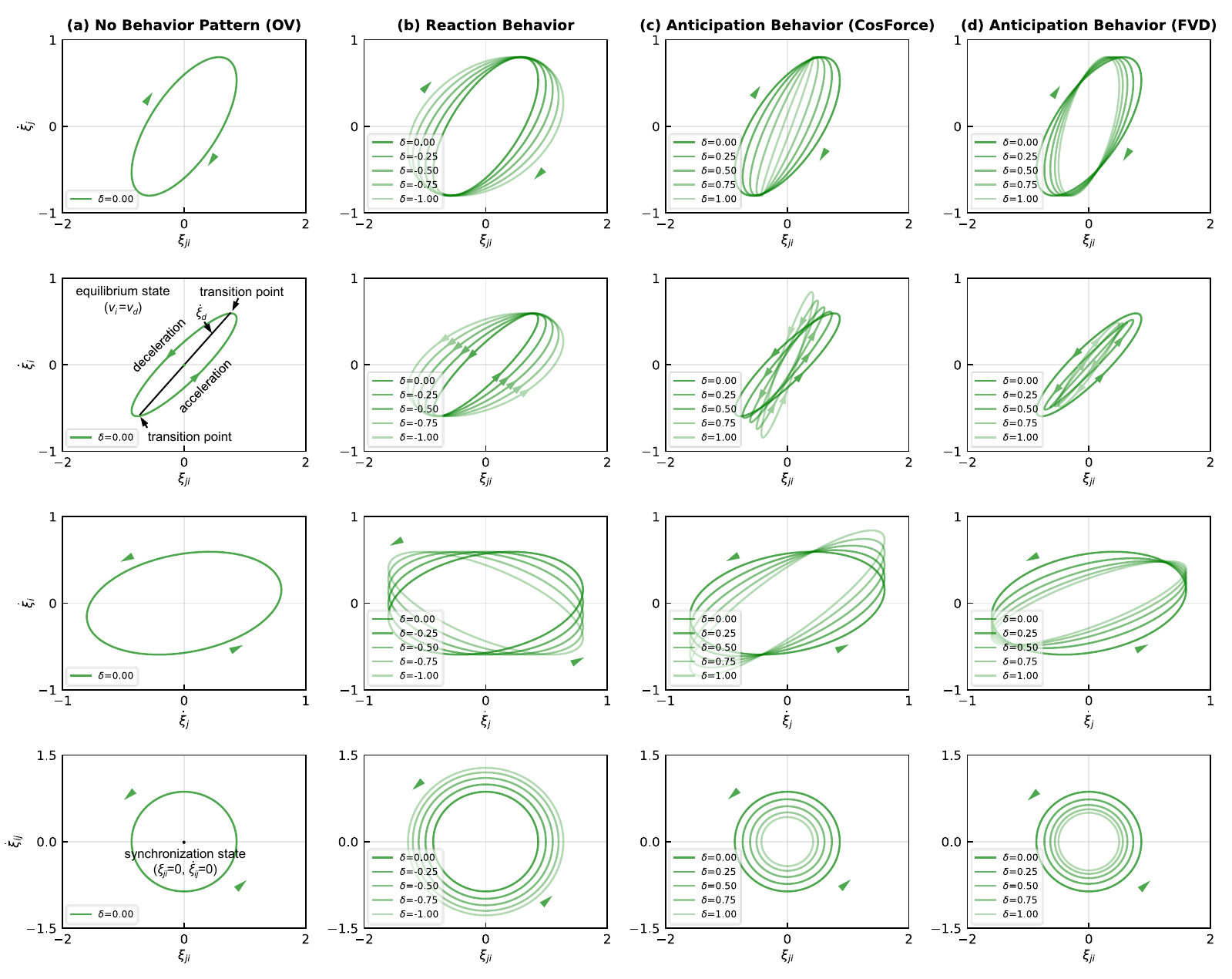}
\caption{Hysteresis characteristics in the four models. Parameter settings for the input wave function: $A = 0.8, \omega = 1$.}
\label{fig3}
\end{figure}

\begin{figure}[ht!]
\nolinenumbers
\centering
\includegraphics[scale=0.65]{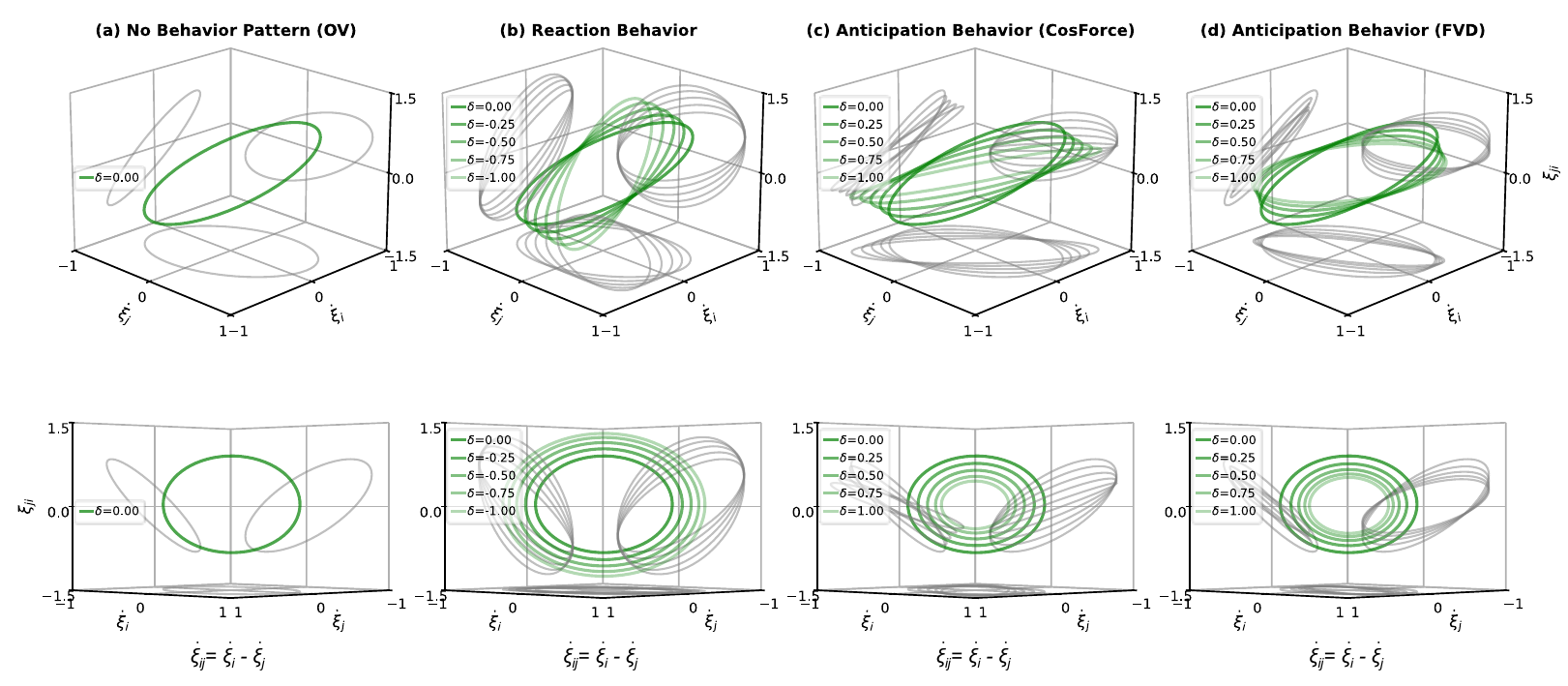}
\caption{3D hysteresis of the four models.}
\label{fig4}
\end{figure}

We have obtained the periodic wave functions describing traffic oscillations of state sets $(\dot{{\xi}}_{j},\dot{{\xi}}_{i}, {\xi}_{ji}, \dot{\xi}_{ij})$, with their corresponding amplitudes and phases clearly defined. By extending the wave functions to the complex plane, it becomes evident that the hysteresis loop originates from the synthesis of two orthogonal waves with different phases. Since these waves share an identical frequency, the resulting composite wave forms an elliptical shape in the phase plane. Within the scope of a single period, the phase lag of the imaginary axis relative to the real axis is defined as $\phi_{\Delta} \in(-\pi,\pi)$. In this condition, the evolution direction of the hysteresis is uniquely determined by the phase lag: when $\phi_{\Delta}<0$, the hysteresis trajectory evolves clockwise; when $\phi_{\Delta}=0$, the hysteresis degenerates into a straight line passing through the origin; and when $\phi_{\Delta}>0$, the trajectory evolves counterclockwise.

\subsubsection{Hysteresis of $\dot{{\xi}}_j$ and ${\xi}_{ji}$}

Let $\theta = \omega t$, the leader speed oscillation and the relative distance oscillation are given by:
\begin{equation}
    \dot{{\xi}}_j(t) = A\omega \cos\theta,
    \label{eq:vj_new}
\end{equation}

\begin{equation}
    {\xi}_{ji}(t) = A\Bigl[(1 - M\cos\phi)\sin\theta - M\sin\phi\cos\theta\Bigr].
    \label{eq:dij_new}
\end{equation}

In matrix form, we rewrite the relationship between the normalized variables as:
\begin{equation}
    \begin{pmatrix}
        \frac{{\xi}_{ji}}{A} \\[2pt]
        \frac{\dot{{\xi}}_j}{A\omega}
    \end{pmatrix}
    =
    \begin{pmatrix}
        1 - M\cos\phi & -M\sin\phi \\[2pt]
        0 & 1
    \end{pmatrix}
    \begin{pmatrix}
        \sin\theta \\[2pt]
        \cos\theta
    \end{pmatrix}.
\end{equation}

Using the identity $\sin^2\theta + \cos^2\theta = 1$, we obtain the following quadratic relation:

\begin{equation}
    \omega^2 \xi_{ji}^2 
    + \left[ (1 - M\cos\phi)^2 + M^2\sin^2\phi \right] \dot{\xi}_j^2 
    + 2M\omega\sin\phi \cdot \xi_{ji}\dot{\xi}_j 
    = A^2\omega^2 (1 - M\cos\phi)^2.
    \label{eq:ellipse_vj_dij}
\end{equation}

For $M \cos\phi \neq 1$, Eq.~\eqref{eq:ellipse_vj_dij} represents an ellipse in the $(\xi_{ji}, \dot{\xi}_j)$ plane, characterizing the hysteresis of the relative distance oscillation versus to the leader's speed oscillation. Correspondingly, the hysteresis loop of relative distance $d_{ji}$ and leader speed $v_j$ is obtained by translating it to the synchronization position in the $(\xi_{ji}, \dot{\xi}_j)$ plane. For the simplicity, we maintain the hysteresis function of the oscillation form, which remains consistent in the subsequent sections.

\subsubsection{Hysteresis of ${\xi}_{ji}$ and $\dot{{\xi}}_i$}

The relative distance and follower speed oscillation are
\begin{equation}
    {\xi}_{ji}(t) = A\sin\theta - A M\sin(\theta + \phi) = A\Bigl[(1 - M\cos\phi)\sin\theta - M\sin\phi\cos\theta\Bigr],
    \label{eq:dij}
\end{equation}

\begin{equation}
    \dot{{\xi}}_i(t) = A M \omega\cos(\theta + \phi) = A M \omega\bigl(\cos\theta\cos\phi - \sin\theta\sin\phi\bigr).
    \label{eq:vi}
\end{equation}

In matrix form, we rewrite the system as:
\begin{equation}
    \begin{pmatrix}
        \frac{{\xi}_{ji}}{A} \\[2pt]
        \frac{\dot{{\xi}}_i}{A\omega}
    \end{pmatrix}
    =
    \begin{pmatrix}
        1 - M\cos\phi & -M\sin\phi \\[2pt]
        -M\sin\phi & M\cos\phi
    \end{pmatrix}
    \begin{pmatrix}
        \sin\theta \\[2pt]
        \cos\theta
    \end{pmatrix}.
\end{equation}

We eliminate the trigonometric terms from the above system and the resulting quadratic relation is:

\begin{equation}
    M^2 \omega^2 \xi_{ji}^2
    + (M^2 - 2M\cos\phi + 1) \dot{\xi}_i^2
    + 2M\sin\phi \cdot \omega \cdot \xi_{ji} \dot{\xi}_i
    = M^2 (M - \cos\phi)^2 A^2 \omega^2.
    \label{eq:ellipse_normalized}
\end{equation}

For $M \neq \cos\phi$, Eq.~\eqref{eq:ellipse_normalized} describes an ellipse in the $({\xi}_{ji}, \dot{\xi}_i)$ plane.

\subsubsection{Hysteresis of $\dot{{\xi}}_j$ and $\dot{{\xi}}_i$}

The follower speed oscillation and leader speed oscillation are
\begin{equation}
    \dot{{\xi}}_i(t) = A\omega\cos\theta,
    \label{eq:vi_original}
\end{equation}
\begin{equation}
    \dot{{\xi}}_j(t) = A M \omega\cos(\theta + \phi) = A M \omega\bigl(\cos\theta\cos\phi - \sin\theta\sin\phi\bigr),
    \label{eq:vj}
\end{equation}

In matrix form, we write the normalized velocities as:
\begin{equation}
    \begin{pmatrix}
        \displaystyle\frac{\dot{{\xi}}_i}{A\omega} \\[8pt]
        \displaystyle\frac{\dot{{\xi}}_j}{A\omega}
    \end{pmatrix}
    =
    \begin{pmatrix}
        1 & 0 \\[2pt]
        M\cos\phi & -M\sin\phi
    \end{pmatrix}
    \begin{pmatrix}
        \cos\theta \\[2pt]
        \sin\theta
    \end{pmatrix}.
\end{equation}

We eliminate $\theta$ from the above system and the resulting quadratic function is:

\begin{equation}
    M^2 \dot{\xi}_i^2 + \dot{\xi}_j^2 - 2M\cos\phi \cdot \dot{\xi}_i \dot{\xi}_j = M^2 A^2 \omega^2 \sin^2\phi
    \label{eq:ellipse_velocity}
\end{equation}

For $\sin\phi \neq 0$, Eq.~\eqref{eq:ellipse_velocity} describes an ellipse in the plane of $(\dot{{\xi}}_i, \dot{{\xi}}_j)$.

\subsubsection{Hysteresis of ${\xi}_{ji}$ and $\dot{{\xi}}_{ij}$}

The relative distance oscillation and relative speed oscillation are defined by

\begin{equation}
    {\xi}_{ji}(t) = {\xi}_j(t) - {\xi}_i(t)
    = A\sin\theta - A M\sin(\theta + \phi),
    \label{eq:xij_rel}
\end{equation}

\begin{equation}
    \dot{{\xi}}_{ij}(t) = \dot{{\xi}}_i(t) - \dot{{\xi}}_j(t)  
    = A M \omega\cos(\theta + \phi) - A\omega\cos\theta,
    \label{eq:dxij_rel}
\end{equation}

Where $\theta = \omega t$, in matrix form, the normalized variables can be written as

\begin{equation}
    \begin{pmatrix}
        \displaystyle\frac{{\xi}_{ji}}{A} \\[8pt]
        \displaystyle\frac{\dot{{\xi}}_{ij}}{A\omega}
    \end{pmatrix}
    =
    \begin{pmatrix}
        1 - M\cos\phi & -M\sin\phi \\[2pt]
        -M\sin\phi     & M\cos\phi - 1 
    \end{pmatrix}
    \begin{pmatrix}
        \sin\theta \\[2pt]
        \cos\theta
    \end{pmatrix}.
    \label{eq:matrix_xij_dxij}
\end{equation}

Eliminating the trigonometric terms yields the following quadratic relation:

\begin{equation}
    \omega^2 {\xi}_{ji}^2 + \dot{\xi}_{ij}^{\,2} = A^2\omega^2\bigl(1 + M^2 - 2M\cos\phi\bigr).
    \label{eq:circle_dimensional}
\end{equation}

Eq.~\eqref{eq:circle_dimensional} describes an ellipse in the plane of $({\xi}_{ji}, \dot{{\xi}}_{ij})$. Since the phase lag $\phi_{\xi_{ji}, \dot{\xi}_{ij}} = \pi/2$, the ellipse function contains no cross-term, and its axes are aligned with the coordinate axes.

Based on the analysis, a comprehensive table is used to summarize the dynamical properties of these four models, as shown in Table~\ref{table1}. It is straightforward that all hystereses are equivalent to Lissajous figures under the condition of equal frequency. In the context of simple harmonic oscillations, the rendezvous points in Lissajous figures are caused only by frequency inconsistency. Based on the properties of the Bode plot, we understand that given a traffic oscillation of general Fourier series, the dynamical system will output different amplitudes and phases based on the wave frequency. In the complex plane, hysteresis can be understood as the synthesis of waves described by two sets of orthogonal Fourier series. Clearly, this may result in hysteresis with multiple rendezvous points, and the particular structure requires a specific discussion.

\begin{sidewaystable}
\nolinenumbers 
\centering
\caption{Comparison of TD and hysteresis characteristics among the four models.}
\label{tab:comparison}
\footnotesize 
\renewcommand{\arraystretch}{3} 
\setlength{\tabcolsep}{2pt} 
\resizebox{\textwidth}{!}{
\begin{tabular}{|c|c|c|c|c|}
\hline
\textbf{Model} & \textbf{No Behavior Pattern (OV)} & \textbf{Reaction Behavior} & \textbf{Anticipation behavior (CosForce)} & \textbf{Anticipation behavior (FVD)} \\
\hline
\textbf{Dynamics (ODE)} & 
$\ddot{x}_i + \frac{\dot{x}_i}{\tau} = \frac{x_j - x_i}{t_h \tau}$ &
$\ddot{x}_i + \frac{\dot{x}_i}{\tau} = \frac{x_{j(t+\delta)} - x_i}{t_h \tau}, \delta < 0$ &
$\ddot{x}_i + \frac{\dot{x}_i}{\tau} = \frac{(x_j + \delta\dot{x}_j) - x_i}{t_h \tau}, \delta > 0$ &
$\ddot{x}_i + \frac{\dot{x}_i}{\tau} = \frac{(x_j - x_i) + \delta(\dot{x}_j - \dot{x}_i)}{t_h \tau}, \delta > 0$ \\

\hline

\shortstack[c]{\textbf{Transfer Function:} \\ $G(s) = {\xi}_i(s)/{\xi}_j(s)$} & 
$G_{\text{ov}}(s) = \dfrac{1}{t_h \tau s^2 + t_h s + 1}$ &
$G_{\text{re}}(s) = \dfrac{e^{\delta s}}{t_h \tau s^2 + t_h s + 1}$ &
$G_{\text{cf}}(s) = \dfrac{1 + \delta s}{t_h \tau s^2 + t_h s + 1}$ &
$G_{\text{fvd}}(s) = \dfrac{1 + \delta s}{t_h \tau s^2 + (t_h + \delta)s + 1}$ \\

\hline
\textbf{Poles: $s_{1,2}$} & 
\multicolumn{3}{c|}{$s_{1,2} = -\dfrac{1}{2\tau} \pm \dfrac{1}{\sqrt{t_h\tau}} \sqrt{\dfrac{t_h}{4\tau} - 1}$} &
$s_{1,2} = \dfrac{-(t_h + \delta) \pm \sqrt{(t_h + \delta)^2 - 4 t_h \tau}}{2 t_h \tau}$ \\

\hline
\textbf{Input: $\xi_j(t)$} & \multicolumn{4}{c|}{$A\sin(\omega t)$} \\

\hline
\shortstack[c]{\textbf{Output}\\[2pt]
\textbf{(Steady-state):}\\[2pt]
$\xi_i(t)$} & 
\shortstack[c]{$A M_{\text{ov}}\sin(\omega t + \phi_{\text{ov}}),$ \\[8pt]
$M_{\text{ov}} = \dfrac{1}{\sqrt{(1 - t_h\tau\omega^2)^2 + (t_h\omega)^2}},$ \\[8pt]
$\phi_{\text{ov}} = - \arg\bigl(1 - t_h\tau\omega^2 + j t_h\omega\bigr)$} &
\shortstack[c]{$A M_{\text{re}}\sin(\omega t + \phi_{\text{re}}),$ \\[8pt]
$M_{\text{re}} = \dfrac{1}{\sqrt{(1 - t_h\tau\omega^2)^2 + (t_h\omega)^2}},$ \\[8pt]
$\phi_{\text{re}} = \delta\omega - \arg\bigl(1 - t_h\tau\omega^2 + j t_h\omega\bigr)$} &
\shortstack[c]{$A M_{\text{cf}}\sin(\omega t + \phi_{\text{cf}}),$ \\[8pt]
$M_{\text{cf}} = \dfrac{\sqrt{1 + (\delta\omega)^2}}{\sqrt{(1 - t_h\tau\omega^2)^2 + (t_h\omega)^2}},$ \\[8pt]
$\phi_{\text{cf}} = \arg\bigl(1 + j\,\delta\omega\bigr)
 - \arg\bigl(1 - t_h\tau\omega^2 + j\,t_h\omega\bigr)$} &
\shortstack[c]{$A M_{\text{fvd}}\sin(\omega t + \phi_{\text{fvd}}),$ \\[8pt]
$M_{\text{fvd}} = \dfrac{\sqrt{1 + (\delta\omega)^2}}{\sqrt{(1 - t_h\tau\omega^2)^2 + ((t_h + \delta)\omega)^2}},$ \\[8pt]
$\phi_{\text{fvd}} = \arg\bigl(1 + j\,\delta\omega\bigr)
 - \arg\bigl(1 - t_h\tau\omega^2 + j\,(t_h + \delta )\omega\bigr)$} \\
\hline

\shortstack[c]{\textbf{Time Delay:} \\ $TD_{\dot{\xi}_j, \xi_{ji}}=\phi_{\dot{\xi}_j, \xi_{ji}}/\omega$} & \multicolumn{4}{c|}{%
$\dfrac{\frac{\pi}{2} - \arg(1 - Me^{j\phi})}{\omega}$} \\
\hline

\shortstack[c]{\textbf{Time Delay:} \\ $TD_{\xi_{ji}, \dot{\xi}_i}=\phi_{\xi_{ji}, \dot{\xi}_i}/\omega$}  & \multicolumn{4}{c|}{%
$\dfrac{\arg(1 - Me^{j\phi}) - \phi - \frac{\pi}{2}}{\omega}$} \\
\hline

\shortstack[c]{\textbf{Time Delay:} \\ $TD_{\dot{\xi}_j, \dot{\xi}_i}=\phi_{\dot{\xi}_j, \dot{\xi}_i}/\omega$}  & \multicolumn{4}{c|}{%
$-\phi/{\omega}$} \\
\hline

\shortstack[c]{\textbf{Time Delay:} \\ $TD_{\xi_{ji}, \dot{\xi}_{ij}}=\phi_{\xi_{ji}, \dot{\xi}_{ij}}/\omega$} & \multicolumn{4}{c|}{%
${\pi}/{2\omega}$} \\
\hline

\textbf{Hysteresis: $E_{\dot{\xi}_j, \xi_{ji}}$} & \multicolumn{4}{c|}{%
$\omega^2 \xi_{ji}^2 
 + \left[ (1 - M\cos\phi)^2 + M^2\sin^2\phi \right] \dot{\xi}_j^2 
 + 2M\omega\sin\phi \cdot \xi_{ji}\dot{\xi}_j 
= A^2\omega^2 (1 - M\cos\phi)^2,\ \xi_{ji}=\xi_{j}-\xi_{i}$} \\
\hline

\textbf{Hysteresis: $E_{\xi_{ji}, \dot{\xi}_i}$} & \multicolumn{4}{c|}{%
$M^2 \xi_{ji}^2 \omega^2
+ (M^2 - 2M\cos\phi + 1) \dot{\xi}_i^2
+ 2M\sin\phi \cdot \omega \cdot \xi_{ji} \dot{\xi}_i
= M^2 (M - \cos\phi)^2 A^2 \omega^2,\ \xi_{ji}=\xi_{j}-\xi_{i}$} \\

\hline
\textbf{Hysteresis: $E_{\dot{\xi}_{j}, \dot{\xi}_i}$} & \multicolumn{4}{c|}{%
$M^2 \dot{\xi}_i^2 + \dot{\xi}_j^2 - 2M\cos\phi \cdot \dot{\xi}_i \dot{\xi}_j = M^2 A^2 \omega^2 \sin^2\phi$} \\

\hline
\textbf{Hysteresis: $E_{\xi_{ji},\dot{\xi}_{ij}}$} & \multicolumn{4}{c|}{%
$\omega^2 {\xi}_{ji}^2 + \dot{\xi}_{ij}^{\,2} = A^2\omega^2\bigl(1 + M^2 - 2M\cos\phi\bigr),\ \xi_{ji}=\xi_{j}-\xi_{i}, \dot{\xi}_{ij}=\dot{\xi}_{i}-\dot{\xi}_{j}$} \\
\hline
\end{tabular}
}
\label{table1}
\end{sidewaystable}

In the 3D phase space $(\dot{{\xi}}_{j},\dot{{\xi}}_{i}, {\xi}_{ji})$, the evolution trajectory of the dynamical system is a 3D ellipse synthesized from three orthogonal wave. The projections of the dynamical trajectory onto different phase planes constitute the hysteresis in a general 2D sense. Specifically, the ellipse projection in the $({\xi}_{ji}, \dot{\xi}_i)$ plane represents the hysteresis in FD as traditionally understood (statistically, linear density is the reciprocal of headway). Given the wave function parameters ($A = 0.8, \omega = 1$), we depict the corresponding hysteresis in different phase planes as shown in Fig.~\ref{fig3}. Fig.~\ref{fig4} illustrates the evolution trajectories in the phase space corresponding to the four models. These 3D ellipses convey the complete dynamical information of the traffic oscillations. The parameter settings for the wave functions here have no specific significance. In single file pedestrian or road traffic contexts, these parameters can be freely replaced while preserving the dynamical properties.

Given the settings for the time scale of anticipation and reaction behaviors, when $\delta=0.5$, the hysteresis of the FVD model in the $({\xi}_{ji}, \dot{{\xi}}_{i})$ plane is a straight line passing through the origin. This depicts a critical equilibrium state of $v_i=v_d$, corresponding to the statistical metric $TD = \phi_{d_{ji}, v_i} / \omega = 0$. When the $TD$ crosses the zero point, the hysteresis flip. Another observation regarding Fig.~\ref{fig3} and Fig.~\ref{fig4} is that the scales of anticipation and reaction affect the geometry of the hysteresis, particularly in the $({\xi}_{ji}, \dot{{\xi}}_{ij})$ plane. It can be observed that anticipation behavior promotes the contraction of the hysteresis, whereas reaction behavior promotes its expansion.

\subsection{Time-To-Collision}

We understand that hysteresis in the $({\xi}_{ji}, \dot{{\xi}}_{i})$ plane represents the dynamical mechanisms of individual deviation from the equilibrium state, which can be directly evaluated through $TD$ statistics. However, hysteresis in the $({\xi}_{ji}, \dot{{\xi}}_{ij})$ plane represents the dynamical mechanisms of system deviation from the synchronization state, exhibiting different properties. Specifically, according to Eq.~\eqref{phi}, it can be understood that $TD_{\xi_{ji}, \dot{\xi}_{ij}}=\pi/2\omega$ remains constant under given input. Therefore, from the perspective of an observer, which quantitative metrics are effective for evaluating hysteresis in the $({\xi}_{ji}, \dot{{\xi}}_{ij})$ plane? That is $TTC$.

The distance $d_{ji}$ is defined by shifting the perturbation terms to the synchronization gap $d$:

\begin{equation}
    d_{ji}(t) = d + \xi_{ji}(t), \quad v_{ij}(t) = \dot{\xi}_{ij}(t)
\end{equation}

Based on Eq.~\eqref{eq:matrix_xij_dxij}, By defining the phase angle $\alpha$ as:
\begin{align}
    \alpha &= \arg\bigl( (1 - M\cos\phi) + j(M\sin\phi) \bigr)
\end{align}

The perturbation components from the matrix form are simplified to:
\begin{align}
    \xi_{ji}(t) &= A \sqrt{1 + M^2 - 2M\cos\phi} \sin(\omega t - \alpha) \\
    \dot{\xi}_{ij}(t) &= -A \sqrt{1 + M^2 - 2M\cos\phi} \omega \cos(\omega t - \alpha)
\end{align}

We derive the complete analytical expression of $TTC$ in time domain:
\begin{equation}
    TTC(t) = \frac{d_{ji}}{v_{ij}} = \frac{d + A \sqrt{1 + M^2 - 2M\cos\phi} \sin(\omega t - \alpha)}{-A  \omega \sqrt{1 + M^2 - 2M\cos\phi} \cos(\omega t - \alpha)}
\end{equation}

The final form is:
\begin{equation}
    TTC(t) = -\frac{1}{\omega} \left[ \frac{d}{A \sqrt{1 + M^2 - 2M\cos\phi}} \sec(\omega t - \alpha) + \tan(\omega t - \alpha) \right].
    \label{eq:TTC_final_full}
\end{equation}

\begin{figure}[ht!]
\nolinenumbers
\centering
\includegraphics[scale=0.4]{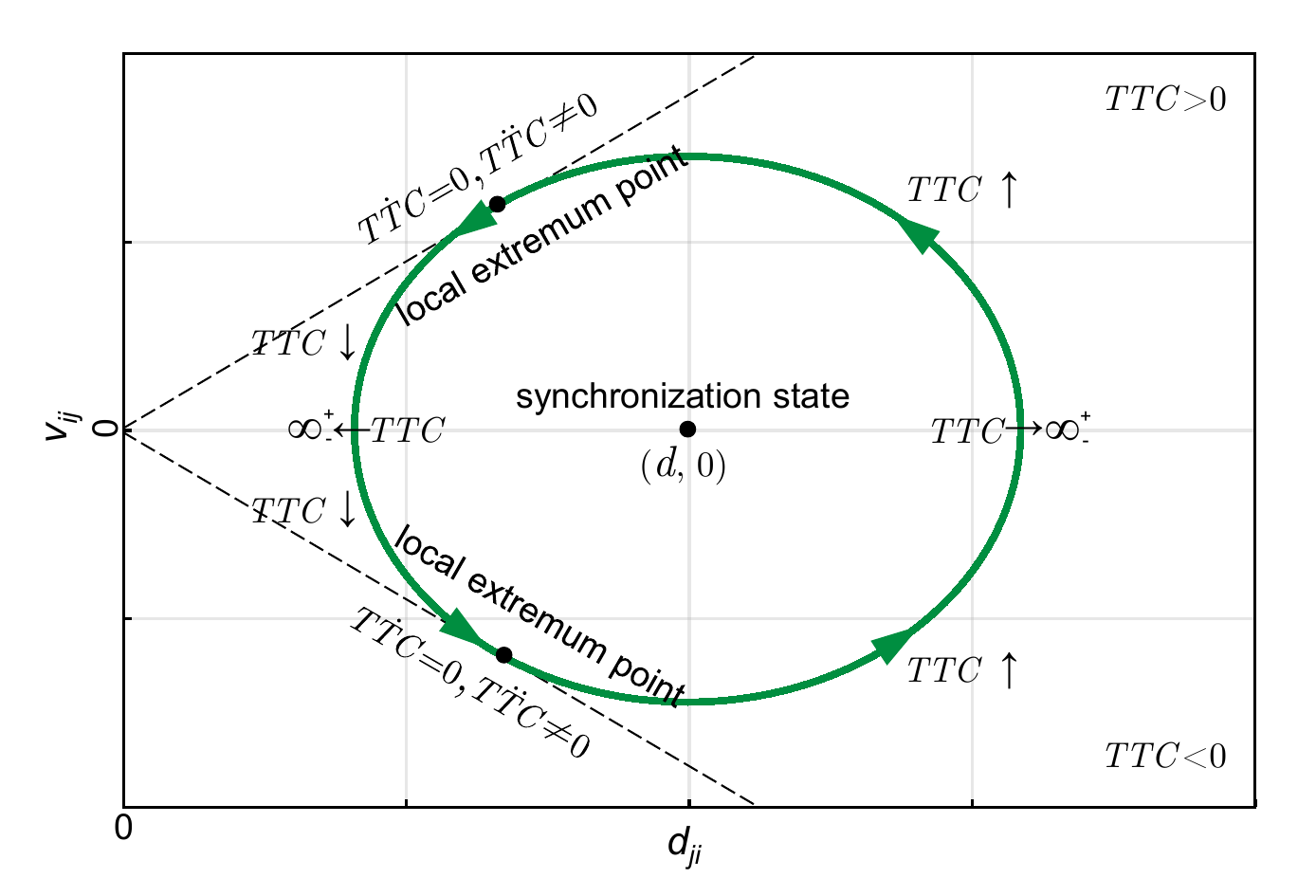}
\caption{Dynamic evolution of $TTC$ in phase trajectory.}
\label{fig5}
\end{figure}

\begin{figure}[ht!]
\nolinenumbers
\centering
\includegraphics[scale=0.65]{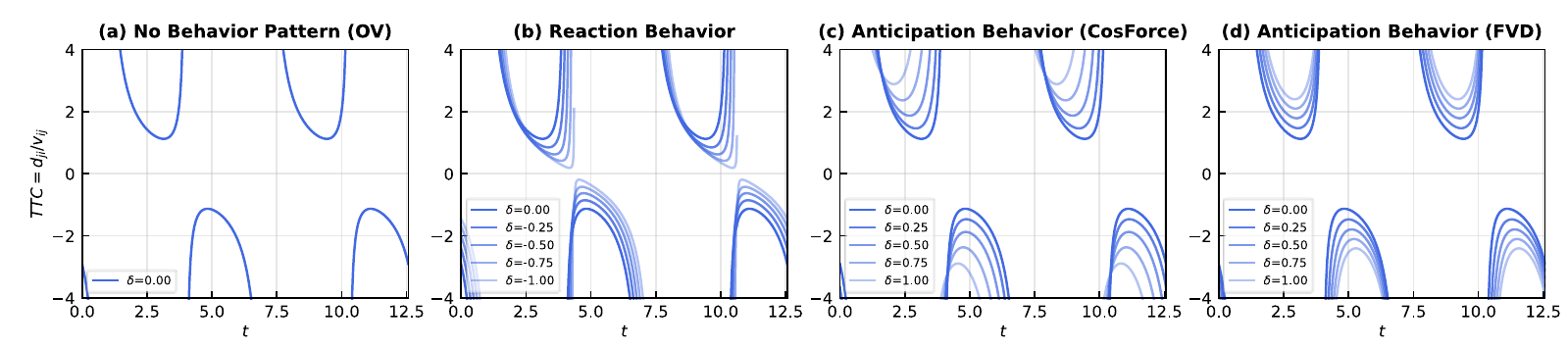}
\caption{Dynamic evolution of $TTC$ in the four models.}
\label{fig6}
\end{figure}

Fig.~\ref{fig5} clearly illustrates the evolution of $TTC$ as the hysteresis evolves in the $({\xi}_{ji}, \dot{{\xi}}_{ij})$ plane. In this absolute coordinate system, the negative slope of the line orthogonal to the ray connecting the hysteresis loop to the origin is equivalent to the $TTC$. The local extreme points of $TTC$ correspond to the tangency points between the rays and the ellipse. We set the synchronized flow headway as 1.3m in pedestrian context. Given the wave function parameters ($A = 0.8, \omega = 1$), the variation of $TTC$ based on Eq.~\ref{eq:TTC_final_full} is shown in Fig.~\ref{fig6}. From the figure, we understand that corresponding to the  expansion and contraction of the hysteresis caused by reaction and anticipation behaviors, the corresponding $TTC$ exhibits the tendency to approach zero (i.e., collision) or away from zero, statistically. Furthermore, $TTC$ can serve not only as an indicator of hysteresis in normal traffic oscillations, but more importantly, it functions to identify extreme dynamics in traffic, which will be discussed in the subsequent.

\section{Numerical experiment}

In this section, based on the analytical results above, we will conduct a numerical experiment on the dynamical properties of the four models. In the numerical results, we shift the oscillation back to the synchronized flow to obtain the absolute coordinates. The synchronized flow location in the phase space is set at $d_{ji} = 1.3 \text{m}$ and $v_{j} = v_{i} = 1 \text{m/s}$, corresponding to a constant time headway of $t_{h} = 1.3 \text{s}$ in pedestrian context. By adjusting these parameters, the results can be transferred to road traffic without altering any dynamical properties.

\begin{figure}[ht!]
\nolinenumbers
\centering
\includegraphics[scale=1.2]{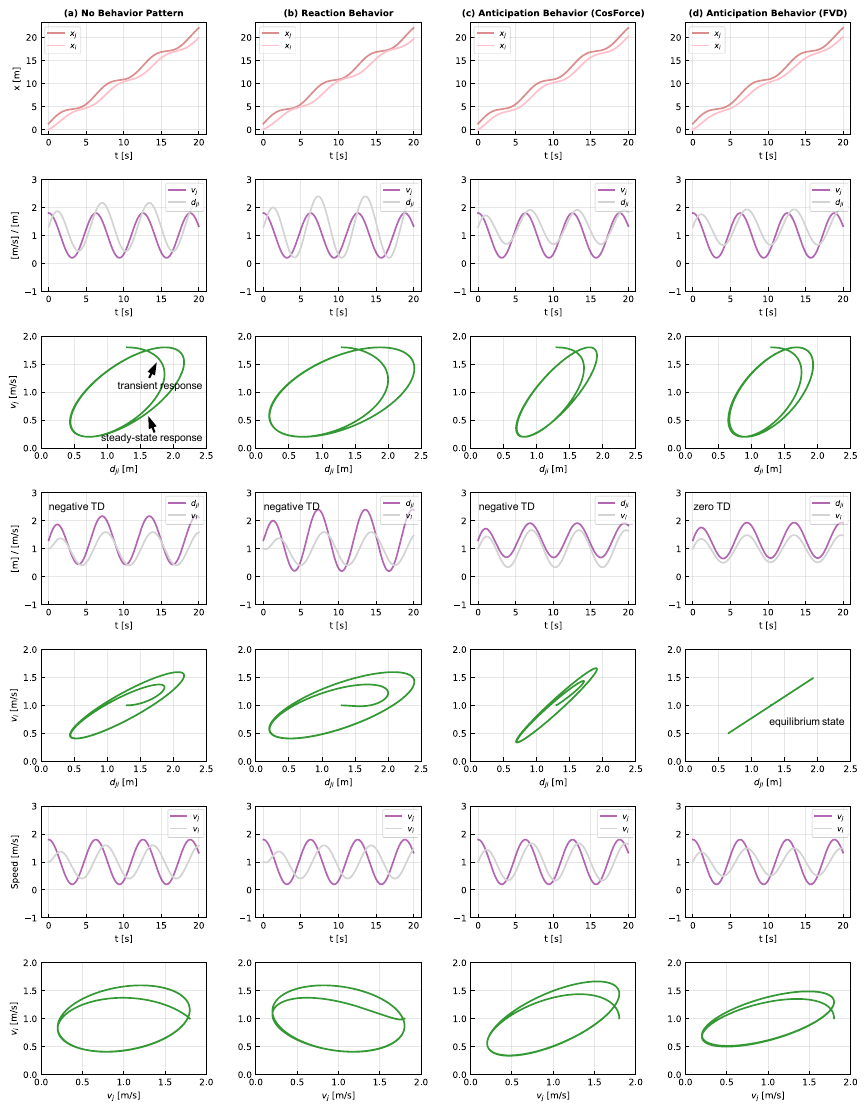}
\caption{Numerical $TD$ and hysteresis results of the four models under sinusoidal input ($A=0.8, \omega=1$).}
\label{fig7}
\end{figure}

\begin{figure}[ht!]
\nolinenumbers
\centering
\includegraphics[scale=1.2]{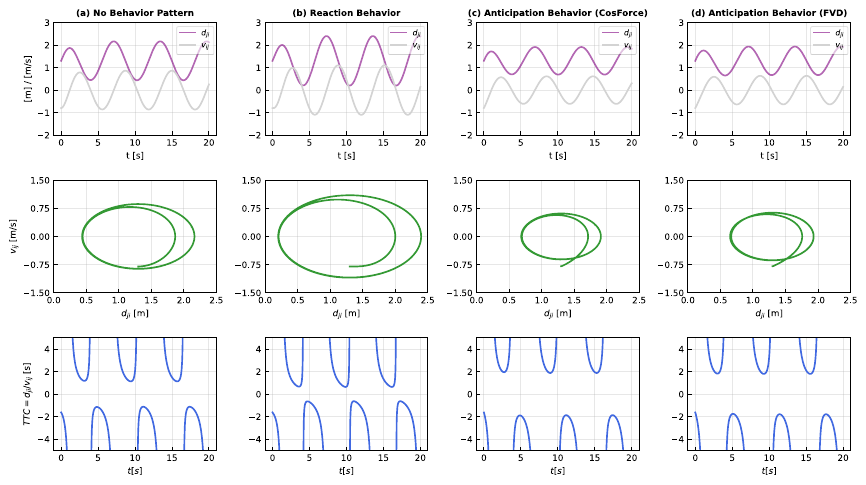}
\caption{Numerical $TTC$ and hysteresis results of the four models under sinusoidal input ($A=0.8, \omega=1$).}
\label{fig8}
\end{figure}

Given the wave function parameters ($A = 0.8, \omega = 1$), $\delta$ is set to -0.5 s in the reaction model and 0.5 s in the anticipation model. We plot the traffic response processes of the four models as shown in Fig.~\ref{fig7}. The figure presents the results of the absolute coordinates in the time domain, where the $TD$ can be directly observed. The numerical results are in agreement with the analytical results. In the FVD model, when $\delta$ is set to 0.5 s, the phases of $d_{ji}$ and $v_i$ are perfectly consistent, corresponding hysteresis in the $({\xi}_{ji}, \dot{{\xi}}_{i})$ plane degenerates into a straight line, which equation is identical to function of $v_d$. At this point, traffic unit $i$ is in a delicate state of equilibrium, i.e., $v_i = v_d$. In Fig.~\ref{fig8}, we plot the curve of the hysteresis in the $({\xi}_{ji}, \dot{{\xi}}_{ij})$ plane along with the corresponding time domain variations of the $TTC$. These results show a consistent trend with the analytical analysis as illustrated above.

\section{Synchronization in Traffic}

\begin{figure}[ht!]
\nolinenumbers
\centering
\includegraphics[scale=0.65]{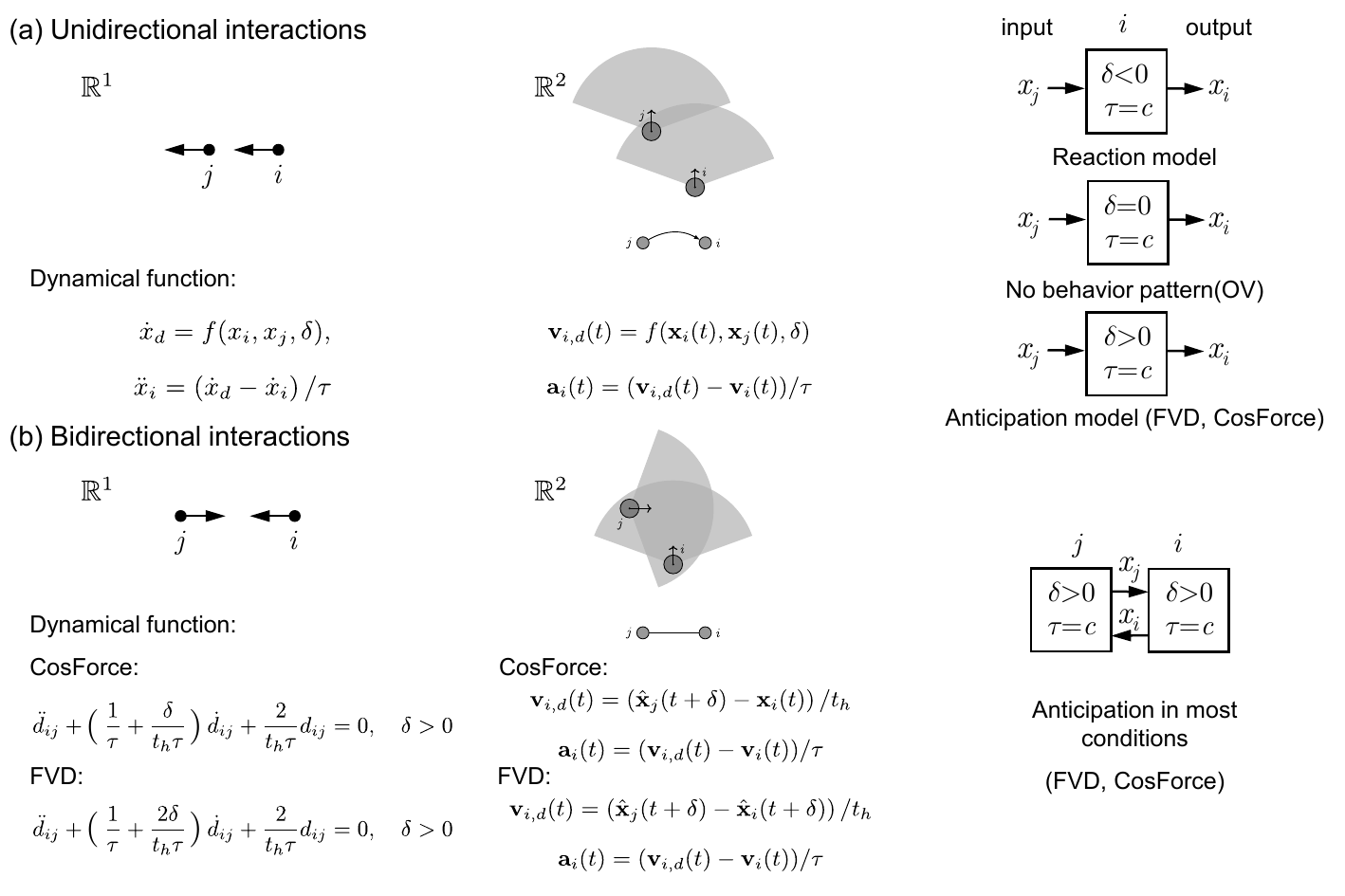}
\caption{Minimized interaction patterns in 1D and 2D spaces.}
\label{fig9}
\end{figure}

\begin{figure}[ht!]
\nolinenumbers
\centering
\includegraphics[scale=0.4]{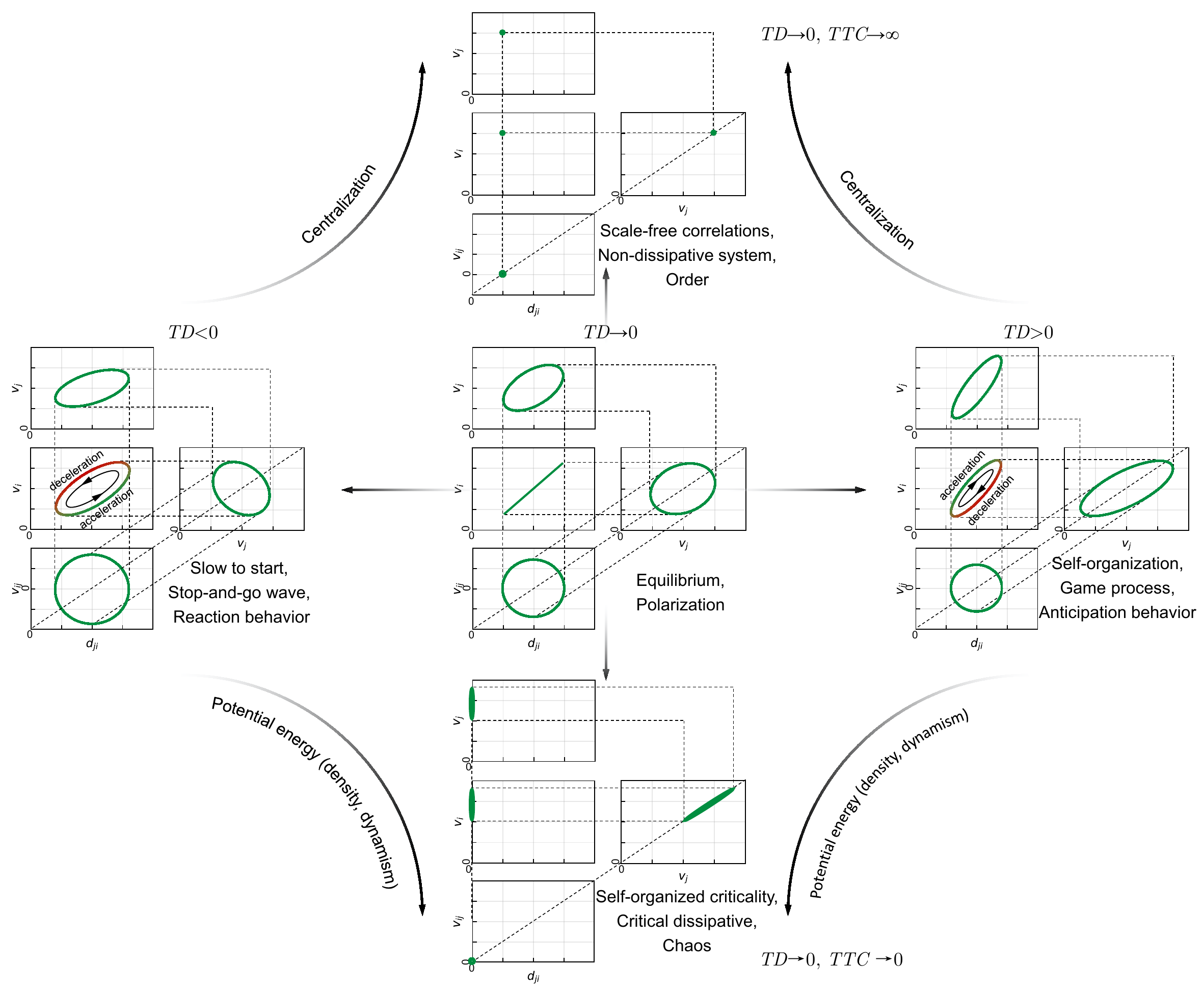}
\caption{Hysteresis characteristics under different dynamical states, characterized by $TD$ and $TTC$.}
\label{fig10}
\end{figure}

\subsection{Kinetic Energy Dissipation Behind Hysteresis}

Building upon the preceding analysis of hysteresis, what fundamental implications can be extracted regarding distinct hysteresis geometry?  From the perspective of energy, the ellipse area in the $({\xi}_{ji}, \dot{{\xi}}_{i})$ plane represents the kinetic energy dissipation $K_{\xi_{ji}, \dot{\xi}_i}$ of traffic unit $i$ as it deviates from individual equilibrium state:

\begin{equation}
K_{\xi_{ji}, \dot{\xi}_i} = \oint\limits_{E_{\xi_{ji}, \dot{\xi}_i}} F_i \, d(\xi_{ji}) = \frac{m}{\tau}\oint\limits_{E_{\xi_{ji}, \dot{\xi}_i}} \left( \dot{\xi}_{d}-\dot{\xi}_i\right) \, d(\xi_{ji})=\frac{m}{\tau} S_{\xi_{ji}, \dot{\xi}_i},
\end{equation}

Corresponding to  Eq.~\eqref{eq:circle_dimensional} and phase lag $\phi_{\xi_{ji}, \dot{\xi}_{i}}$, the area is:

\begin{equation}
S_{\xi_{ji}, \dot{\xi}_i} = \pi A_{\xi_{ji}} A_{\dot{\xi}_i}  |\sin\phi_{{\xi}_{ji}, \dot{{\xi}}_{i}}|.
\end{equation}

From the equilibrium state of individual perspective, the system manifests as the hysteresis with zero area in the phase space of $d_{ji}$ and $v_i$. This signifies the zero kinetic energy dissipation within the individual dynamics, corresponding to the condition $\phi_{{\xi}_{ji}, \dot{{\xi}}_{i}}=0$, i.e., $TD = \phi_{d_{ji}, v_i} / \omega = 0$. Revisiting Fig.~\ref{fig3} reveals that the reaction behavior enhances the kinetic energy dissipation of individuals deviating from the equilibrium state, whereas moderate anticipation behavior weakens such dissipation. However, when anticipation exceeds a critical threshold ($TD = \phi_{d_{ji}, v_i} / \omega = 0$), excessive anticipation leads to an increase in the dissipation magnitude.

Consider an idealized periodic unidirectional flow consisting only of traffic units $i$ and $j$. In this system, $\xi_{ij} = -\xi_{ji}$. The ellipse area in $({\xi}_{ji}, \dot{{\xi}}_{ij})$ plane, illustrating the systematic kinetic energy dissipation $K_{\xi_{ji}, \dot{\xi}_{ij}}$ away from synchronization state within the traffic oscillations, is:

\begin{equation}
K_{\xi_{ji}, \dot{\xi}_{ij}} = \oint\limits_{E_{\xi_{ji}, \dot{\xi}_{ij}}} F_i \, d(\xi_{ji}) + \oint\limits_{E_{\xi_{ji}, \dot{\xi}_{ij}}} F_j \, d(\xi_{ij}) = \frac{m}{\tau}\oint\limits_{E_{\xi_{ji}, \dot{\xi}_{ij}}} \left( \dot{\xi}_{d} - \dot{\xi}_i\right) - \left(\dot{\xi}_{d} - \dot{\xi}_j\right) \, d(\xi_{ji})= \frac{m}{\tau}\oint\limits_{E_{\xi_{ji}, \dot{\xi}_{ij}}} - \dot{\xi}_{ij} \, d(\xi_{ji})=\frac{m}{\tau} S_{\xi_{ji}, \dot{\xi}_{ij}}
\end{equation}

Corresponding to phase lag $\phi_{\xi_{ji}, \dot{\xi}_{ij}}=\pi/2$, the area is:

\begin{equation}
S_{\xi_{ji}, \dot{\xi}_{ij}} = \pi A_{\xi_{ji}} A_{\dot{\xi}_{ij}}|\sin\phi_{{\xi}_{ji}, \dot{{\xi}}_{ij}}|= \pi A_{\xi_{ji}} A_{\dot{\xi}_{ij}}
\end{equation}

Similarly, as illustrated in Fig.~\ref{fig3}, the hysteresis loop in the $({\xi}_{ji}, \dot{{\xi}}_{ij})$ plane undergoes expansion in the reaction model and contraction in the anticipation model. This result indicates that reaction behavior enhances the systematic energy dissipation induced by traffic oscillations, whereas anticipation behavior attenuates it. Corresponding to synchronization state of system perspective, the hysteresis manifests as zero area in the $(d_{ji}, v_{ij})$ plane. In this state, $v_{ij} \to 0$ approaches zero. The two cases where $d_{ji} = 0$ and $d_{ji} \neq 0$ represent the two extreme conditions of dynamical evolution, namely chaos and perfect order. These states correspond to the quantitative metric where $TTC = d_{ji} / v_{ij} \to 0$ and $TTC = d_{ji} / v_{ij} \to \infty$, respectively.

\subsection{Phase Diagram of Synchronization}

In the analysis above, our analytical investigation starting from the simplest 1D particle system. The mechanisms of the hysteresis elegant explained many obscure properties in traffic. Expanding the scope to a more general condition, compared to 1D motion, the characteristics of traffic in 2D space are more diverse and general. In 2D motion with multi-body interactions, even if local evolution can be described using linear differential equations, from a global perspective, the evolution of motion is non linear and unstable. Consequently, for 2D traffic analysis, we must resort to finite difference calculations as a compromise. Nevertheless, in both 1D and 2D spaces, the local interaction rules of traffic response units are convergent. These can be categorized into unidirectional interactions and bidirectional interactions, which constitute the minimized patterns of the fundamental traffic interactions, as shown in Fig.~\ref{fig9}. Based on these properties, we attempt to summarize the characteristics of hysteresis to more general scenarios. By utilizing hysteresis to connect these components, we obtain the phase diagram of synchronization shown in Fig.~\ref{fig10}. Based on the above analysis, the phase diagram is depicted by two concise time metrics, $TD$ and $TTC$, which can derived from direct observations of traffic states.

\begin{figure}[ht!]
\nolinenumbers
\centering
\includegraphics[scale=0.65]{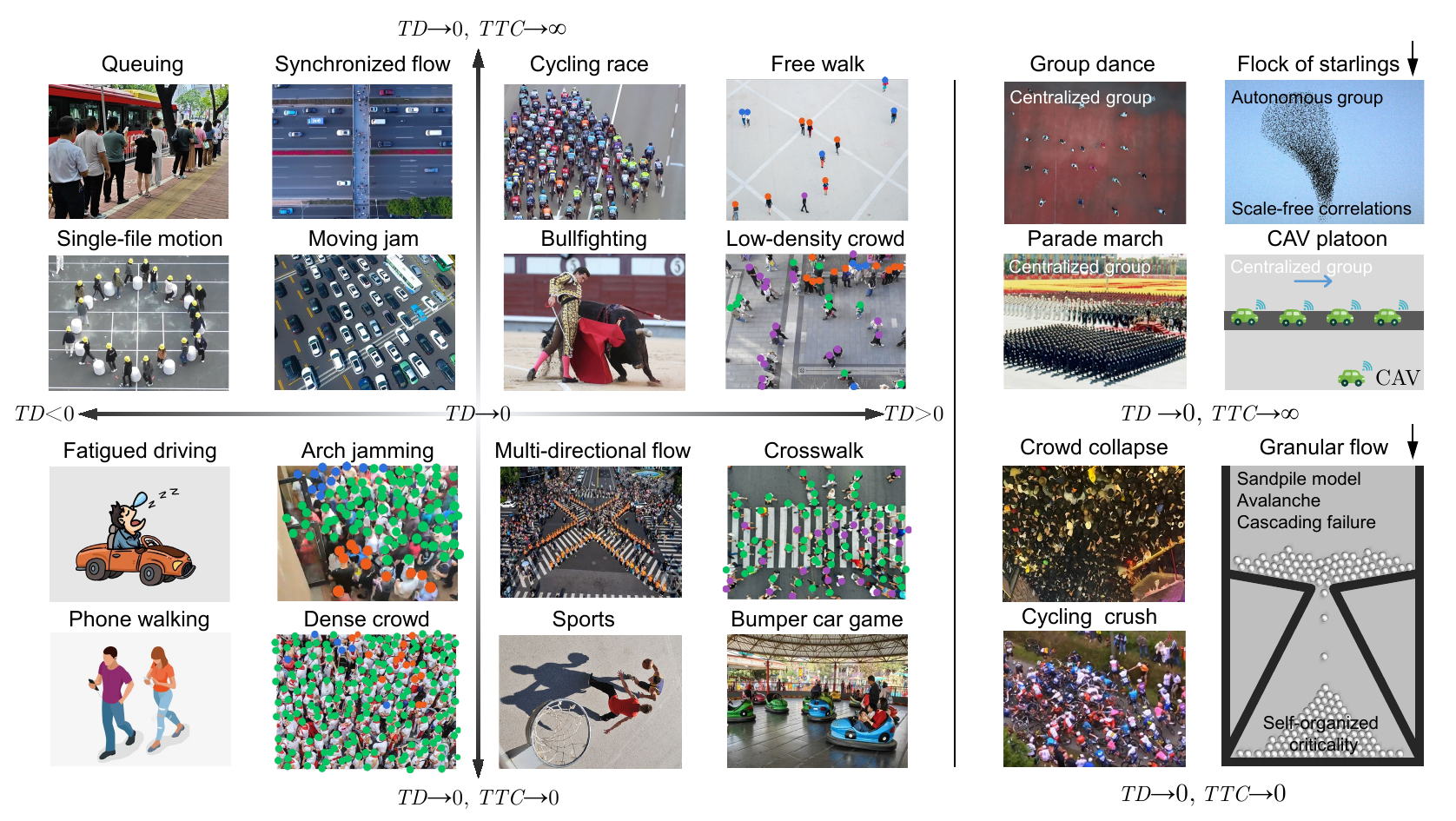}
\caption{Synchronization phase diagrams of real-world traffic based on $TD$ and $TTC$. Arrows indicate non traffic dynamical phenomena that are included due to the consistency of their dynamical characteristics.}
\label{fig11}
\end{figure}

In the phase diagram of synchronization, we connect quantitative hysteresis features with qualitative dynamical phenomena. Most importantly, as $TD$ approaches zero, under different environmental drivers, the dynamical properties of traffic will converge to two extreme conditions: perfect order and complete chaos, as shown in Fig.~\ref{fig10}. To deepen the understanding of this phase diagram, we provide the mapping of the phase diagram of real world traffic, as shown in Fig.~\ref{fig11}. This synchronization phase diagram effectively characterizes various dynamical behaviors in nature, providing deeper insights into complex dynamical systems.

\section{Conclusion}

This paper investigates four types of second-order models in traffic established on the constant time-headway strategy. The parameters of the dynamical system retain clear physical meanings, ensuring that the dimensional consistency adheres to principles of mathematical and physical rigor. We understand that mutually orthogonal wave on different phase axes synthesize the elliptical evolution trajectory of the traffic dynamical system in three dimensional phase space. Essentially, the 2D hystereses are projections of this dynamical trajectory onto different phase planes. Furthermore, we demonstrate how the area of hysteresis quantifies the kinetic energy dissipation stemming from traffic perturbations. From a statistical perspective, two concise time metrics for hysteresis evaluation, $TD$ and $TTC$, are introduced. Finally, a phase diagram of synchronization based on $TD$ and $TTC$ is developed to characterize various dynamical behaviors in traffic. 

So, what is traffic hysteresis? It is the trajectory of dynamic system in phase space; the dissipation of kinetic energy shapes its shadow, while $TD$ and $TTC$ trace its twin footprints.

\centerline{}


\bibliographystyle{aasjournal}

\begin{thebibliography}{}
\expandafter\ifx\csname natexlab\endcsname\relax\def\natexlab#1{#1}\fi
\providecommand{\url}[1]{\href{#1}{#1}}
\providecommand{\dodoi}[1]{doi:~\href{http://doi.org/#1}{\nolinkurl{#1}}}
\providecommand{\doeprint}[1]{\href{http://ascl.net/#1}{\nolinkurl{http://ascl.net/#1}}}
\providecommand{\doarXiv}[1]{\href{https://arxiv.org/abs/#1}{\nolinkurl{https://arxiv.org/abs/#1}}}

\bibitem[{Bando {et~al.}(1995)Bando, Hasebe, Nakayama, Shibata, \& Sugiyama}]{bando1995dynamical}
Bando, M., Hasebe, K., Nakayama, A., Shibata, A., \& Sugiyama, Y. 1995, Physical review E, 51, 1035, \dodoi{10.1103/PhysRevE.51.1035}

\bibitem[{Chen {et~al.}(2012)Chen, Laval, Ahn, \& Zheng}]{chen2012microscopic}
Chen, D., Laval, J.~A., Ahn, S., \& Zheng, Z. 2012, Transportation Research Part B: Methodological, 46, 1440, \dodoi{10.1016/j.trb.2012.07.002}

\bibitem[{Cordes {et~al.}(2023)Cordes, Chraibi, \& Tordeux}]{cordes2023single}
Cordes, J., Chraibi, M., \& Tordeux, A. 2023, Crowd Dynamics, Volume 4: Analytics and Human Factors in Crowd Modeling, 143, \dodoi{10.1007/978-3-031-46359-4_6}

\bibitem[{Gayah \& Daganzo(2011)}]{gayah2011clockwise}
Gayah, V.~V., \& Daganzo, C.~F. 2011, Transportation Research Part B: Methodological, 45, 643, \dodoi{10.1016/j.trb.2010.11.006}

\bibitem[{Jiang {et~al.}(2024)Jiang, Zhou, Wang, \& Ahn}]{jiang2024dynamic}
Jiang, J., Zhou, Y., Wang, X., \& Ahn, S. 2024, Transportation Research Part B: Methodological, 189, 102979, \dodoi{10.1016/j.trb.2024.102979}

\bibitem[{Jiang {et~al.}(2001)Jiang, Wu, \& Zhu}]{jiang2001full}
Jiang, R., Wu, Q., \& Zhu, Z. 2001, Physical Review E, 64, 017101, \dodoi{10.1103/PhysRevE.64.017101}

\bibitem[{Kesting \& Treiber(2008)}]{kesting2008reaction}
Kesting, A., \& Treiber, M. 2008, Computer-Aided Civil and Infrastructure Engineering, 23, 125, \dodoi{10.1111/j.1467-8667.2007.00529.x}

\bibitem[{Laval(2011)}]{laval2011hysteresis}
Laval, J.~A. 2011, Transportation Research Part B: Methodological, 45, 385, \dodoi{10.1016/j.trb.2010.07.006}

\bibitem[{Makridis {et~al.}(2019)Makridis, Mattas, Borio, \& Ciuffo}]{makridis2019estimating}
Makridis, M., Mattas, K., Borio, D., \& Ciuffo, B. 2019, in 2019 6th international conference on models and technologies for intelligent transportation systems (MT-ITS), IEEE, 1--7, \dodoi{10.1109/MTITS.2019.8883341}

\bibitem[{Mattas {et~al.}(2023)Mattas, Albano, Don{\`a}, He, \& Ciuffo}]{mattas2023relationship}
Mattas, K., Albano, G., Don{\`a}, R., He, Y., \& Ciuffo, B. 2023, Transportation research part B: methodological, 174, 102785, \dodoi{10.1016/j.trb.2023.102785}

\bibitem[{Ni(2025)}]{ni2025there}
Ni, D. 2025, Transportation Research Part B: Methodological, 195, 103206, \dodoi{10.1016/j.trb.2025.103206}

\bibitem[{Saifuzzaman {et~al.}(2017)Saifuzzaman, Zheng, Haque, \& Washington}]{saifuzzaman2017understanding}
Saifuzzaman, M., Zheng, Z., Haque, M.~M., \& Washington, S. 2017, Transportation Research Part B: Methodological, 105, 523, \dodoi{10.1016/j.trb.2017.09.023}

\bibitem[{Treiber {et~al.}(2006)Treiber, Kesting, \& Helbing}]{treiber2006delays}
Treiber, M., Kesting, A., \& Helbing, D. 2006, Physica A: Statistical Mechanics and its Applications, 360, 71, \dodoi{10.1016/j.physa.2005.05.001}

\bibitem[{Treiterer \& Myers(1974)}]{Treiterer}
Treiterer, J., \& Myers, J. 1974, Transportation and traffic theory, 6, 13

\bibitem[{Zhang(1999)}]{zhang1999mathematical}
Zhang, H.~M. 1999, Transportation Research Part B: Methodological, 33, 1, \dodoi{10.1016/S0191-2615(98)00022-8}

\end{thebibliography}

\end{document}